\documentclass[10pt, journal, final, twocolumn]{IEEEtran} 
\usepackage{amsmath,amsfonts}
\usepackage{algorithmic}
\usepackage{algorithm}
\usepackage{array}
\usepackage[caption=false,font=footnotesize,labelfont=sf,textfont=sf]{subfig}
\usepackage{textcomp}
\usepackage{url}
\usepackage{verbatim}
\usepackage{graphicx}
\usepackage{cite}
\usepackage{booktabs}
\usepackage{multirow}

\usepackage{authblk}

\usepackage[justification=centering,labelsep=period]{caption}

\title{Audio Outperforms Text for Visual Decoding}
\author[1,]{Zhengdi Zhang}
\author[1]{Hao Zhang} 
\author[1,*]{Wenjun Xia}

\affil[1]{School of Mathematical Sciences, Jiangsu University}
\affil[*]{Corresponding author: vvjqnwork@outlook.com}                           

\begin{document}
\maketitle
\begin{abstract}

Decoding visual semantic representations from human brain activity is a significant challenge. While recent zero-shot decoding approaches have improved performance by leveraging aligned image-text datasets, they overlook a fundamental aspect of human cognition: semantic understanding is inherently anchored in the auditory modality of speech, not text. To address this, our study introduces the first comparative framework for evaluating auditory versus textual semantic modalities in zero-shot visual neural decoding. We propose a novel brain-visual-auditory multimodal alignment model that directly utilizes auditory representations to encapsulate semantics, serving as a substitute for traditional textual descriptors. Our experimental results demonstrate that the auditory modality not only surpasses the textual modality in decoding accuracy but also achieves higher computational efficiency. These findings indicate that auditory semantic representations are more closely aligned with neural activity patterns during visual processing. This work reveals the critical and previously underestimated role of auditory semantics in decoding visual cognition and provides new insights for developing brain-computer interfaces that are more congruent with natural human cognitive mechanisms.
	
\end{abstract}
\section{Introduction}

Decoding visual neural representations in the human brain is a central topic in understanding human cognition and developing intelligent brain-computer systems. By analyzing brain activity patterns, researchers can explore how the brain transforms external stimuli into internal representations \cite{Haxby2001, Kamitani2005}, providing a scientific foundation for uncovering visual processing mechanisms and developing brain-inspired intelligent machines. This research also lays a crucial groundwork for the advancement of brain-computer interfaces (BCIs) \cite{Wolpaw2002}, enabling information transfer between humans and machines to gradually move from the signal level to the semantic level.

Over the past decade, significant progress has been made in the field of neural decoding. Researchers have proposed a variety of methods based on functional magnetic resonance imaging (fMRI) \cite{Naselaris2011}, electroencephalography (EEG) \cite{Horikawa2017}, and deep learning \cite{LeCun2015} to reconstruct visual stimuli or semantic concepts from neural signals. Among these, zero-shot neural decoding (ZSND) has emerged as a research focus in recent years \cite{Palatucci2009}, aiming to predict the semantic information corresponding to the subject's brain activity for unseen categories. Existing studies primarily rely on two types of knowledge: visual semantic knowledge derived from the visual feature space (e.g., CNN features \cite{Krizhevsky2012}) and linguistic semantic knowledge derived from language embeddings (e.g., word2vec \cite{Mikolov2013}, BERT \cite{Devlin2019}). These two forms of knowledge jointly constitute the foundation of the interaction between vision and language in the human cognitive system.

\begin{figure*}[!t]
	\centering
	\includegraphics[width=0.4\textwidth]{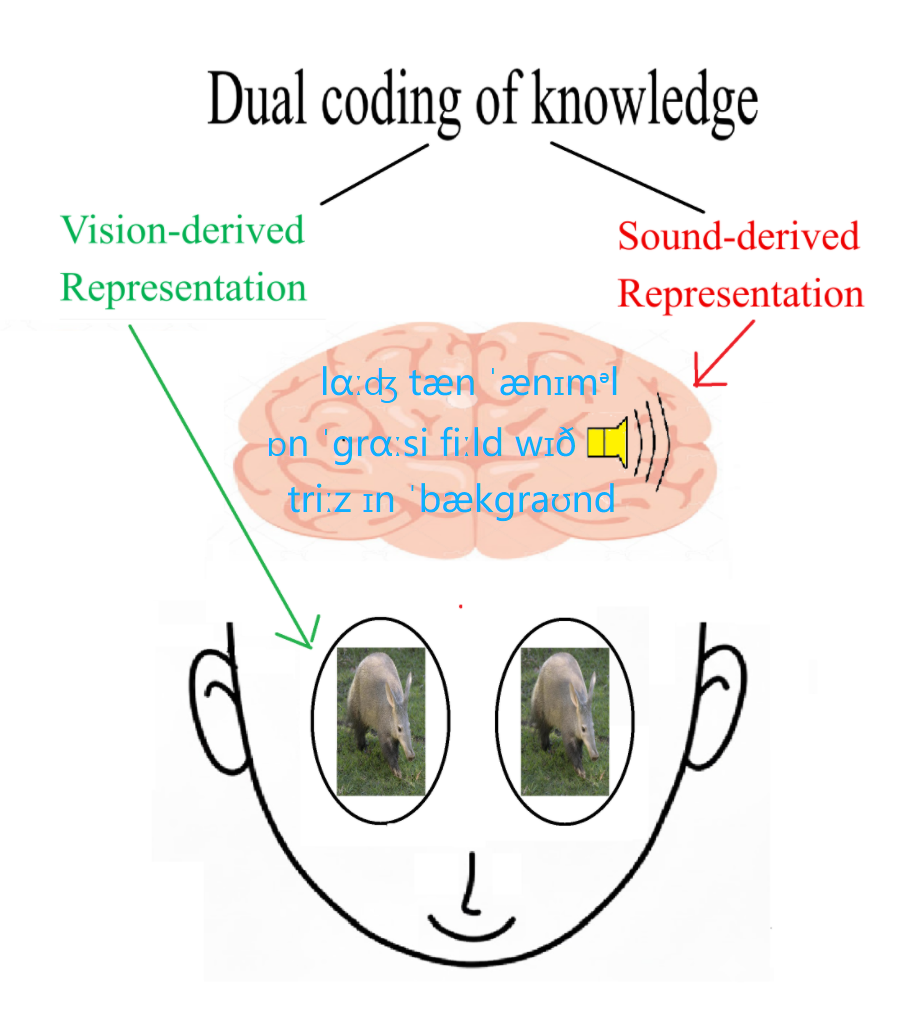} 
	\caption{The proposed brain-visual-audio multimodal alignment framework.}
	\label{Fig:1}
\end{figure*}

In cognitive neuroscience, Dual Coding Theory \cite{paivio1971imagery,clark1991dual} proposes that the neural representation of human concepts relies on the coordinated encoding of vision and language: the visual system provides pictorial representations of perceptual experience, while the language system supplies symbolic and abstract semantic support. Inspired by this theory, recent research has attempted to model this dual-representation mechanism through multimodal learning \cite{baltruvsaitis2018multimodal}. Figure~\ref{Fig:1} depicts the underlying cognitive process, illustrating that knowledge is composed of both visual and linguistic representations. For instance, the BraVL model \cite{Du2023BraVL} integrates three modalities-brain, vision, and language, demonstrating the critical role of joint visual and linguistic representation in neural semantic decoding.

However, existing approaches still face key limitations. Although the introduction of the language modality significantly improves decoding performance, most studies rely solely on “textual language representations,” such as category names or sentence descriptions. While symbolic text captures abstract semantics, it diverges significantly from the auditory representations through which the human brain naturally processes language. In fact, human language comprehension depends not only on written text but primarily on auditory input-speech which involves the multi-level integration of acoustic features, prosodic structure, and semantic information \cite{Hickok2007}. Consequently, text-based brain-language models exhibit limited cognitive plausibility, constraining their ecological validity and neural alignment. To more accurately reflect this cognitive foundation, Figure~\ref{Fig:1} highlights the use of a sound-derived linguistic representation, rather than a text-based one, as the core semantic input to the brain-language model.

From a neurobiological standpoint, auditory language representations offer a more natural pathway for semantic access than written text. The dual-stream model of speech processing \cite{Hickok2007} suggests that the ventral auditory stream
 is specialized for directly mapping acoustic speech signals onto semantic representations, bypassing the intermediate symbolic transformations required for reading. This evolutionary optimization implies that spoken language may align more efficiently with the brain’s semantic system. Moreover, grounded cognition theory \cite{Barsalou2008} argues that conceptual knowledge is rooted in perceptual and sensorimotor experiences rather than abstract symbols. Auditory speech preserves rich prosodic and temporal dynamics, yielding perceptually grounded representations that are theoretically more compatible with time-resolved neural signals such as EEG.

Motivated by these cognitive and neurobiological considerations, this paper proposes a multimodal neural decoding framework based on the sound modality, investigating whether auditory language representations can achieve more effective alignment with human neural activity than textual representations. Specifically, we replace the language modality in the BraVL framework \cite{Du2023BraVL} with an audio modality and construct a trimodal joint model of \textit{brain–vision–sound}. Through multimodal contrastive learning and latent semantic alignment, this framework achieves cross-modal mapping from brain signals to natural semantics, thereby more accurately capturing the neural dynamics of the human brain during language comprehension.

The main contributions of this paper are as follows:

\begin{enumerate} \item Systematic experiments across multiple subjects and category settings demonstrate that the sound modality exhibits higher neural alignment and generalization capabilities than text in brain-signal-based semantic decoding. \item From a cognitive perspective, our study reveals the potential role of auditory representation in visual semantic decoding, underscoring that the auditory dimension of natural language plays an essential role in human semantic processing. This provides new insights for developing multimodal models that are more consistent with human brain function. 
\end{enumerate}

\section{Methods}
Our framework is built upon the BraVL \cite{Du2023BraVL} model. We adopt its VAE architecture, MoPoE fusion (Section 2.2), and MI regularization strategy (Section 2.3). Our primary contribution is the novel replacement of the text modality ($x_t$) with a perceptually-grounded audio modality ($x_a$), as described in Section 2.5."

\subsection{Variational Autoencoder (VAE) Basics}

Following the framework introduced in BraVL \cite{Du2023BraVL}, we build our model upon the standard Variational Autoencoder (VAE) formulation. For a single modality, the objective is to maximize the Evidence Lower Bound (ELBO):
\begin{equation}
	\log p_{\theta}(x) \ge \mathbb{E}_{q_{\phi}(z|x)}[\log p_{\theta}(x|z)] - D_{KL}\!\left[q_{\phi}(z|x)\,\|\,p(z)\right]
\end{equation}
where $q_{\phi}(z|x)$ is the encoder and $p_{\theta}(x|z)$ is the decoder. To enable gradient-based optimization through random sampling, the reparameterization trick is used:
\begin{equation}
	z = \mu + \sigma \times \epsilon, \quad \text{where } \epsilon \sim \mathcal{N}(0, 1)
\end{equation}

\subsection{Multimodal Joint Modeling with MoPoE}
To fuse information from brain ($x_b$), visual ($x_v$), and linguistic ($x_t$) modalities, we adopt the Mixture-of-Products-of-Experts (MoPoE) formulation proposed in BraVL \cite{Du2023BraVL} to define the joint posterior distribution $q_{\Phi}(z|\mathbb{X})$:
\begin{equation}
	q_{\Phi}(z|\mathbb{X}) = \frac{1}{|\mathcal{P}(\mathbb{X})|}\sum_{\mathbb{X}_{s}\in\mathcal{P}(\mathbb{X})}\prod_{x_{m}\in\mathbb{X}_{s}} q_{\Phi_{m}}(z|x_{m})
\end{equation}
where $\mathcal{P}(\mathbb{X})$ is the power set of available modalities $\mathbb{X}$. This allows the model to robustly handle any subset of modalities.

The core generative objective is the multimodal ELBO, $\mathcal{L}_{M}$, which incorporates this joint posterior:
\begin{equation}
	\begin{aligned}
		\mathcal{L}_{M}(\Theta,\Phi) &= \mathbb{E}_{q_{\Phi}(z|\mathbb{X})}\left[\sum_{x_{m}\in\mathbb{X}}\log p_{\theta_{m}}(x_{m}|z)\right] \\
		&\quad - D_{KL}\!\left[\frac{1}{|\mathcal{P}(\mathbb{X})|}\sum_{\mathbb{X}_{s}\in\mathcal{P}(\mathbb{X})}\prod_{x_{m}\in\mathbb{X}_{s}} q_{\Phi_{m}}(z|x_{m})\,\Big\|\,p(z)\right]
	\end{aligned}
\end{equation}

\subsection{Mutual Information Regularization}
We follow the mutual information regularization strategy of BraVL \cite{Du2023BraVL}, introducing intra- and inter-modality objectives to encourage latent coherence and cross-modal alignment.

\subsubsection{Intra-Modality MI Maximization}
To prevent posterior collapse, we maximize the MI between the latent code $z$ and each individual modality $x_m$. This is achieved by maximizing its variational lower bound:
\begin{equation}
	I(z;x_{m}) \ge \mathbb{E}_{z\sim q_{\Phi}(z|\mathbb{X}),x_{m}\sim p_{\theta,m}(x_{m}|z)}[\log Q_{\psi_{m}}(z|x_{m})]+H(z)
\end{equation}
The final regularizer, $\mathcal{L}_{intra}$, sums this objective over all available modalities:
\begin{equation}
	\begin{aligned}
		\mathcal{L}_{intra}(\Theta,\Phi,\Psi) &= \sum_{m}\mathbb{E}_{z\sim q_{\Phi}(z|\mathbb{X}),x_{m}\sim p_{\theta,m}(x_{m}|z)} \\
		&\quad \times \left[\log Q_{\psi_{m}}(z|x_{m})\right]+3H(z)
	\end{aligned}
\end{equation}
\subsubsection{Inter-Modality MI Maximization}
To align the representations from different modalities, we maximize the MI across modalities, similar to the approach in BraVL \cite{Du2023BraVL}. For the trimodal case, the objective is:
\begin{equation}
	\mathcal{L}_{inter}(\Theta,\Phi) = I(x_{b},x_{v};x_{t})+I(x_{b};x_{v},x_{t})+I(x_{b},x_{t};x_{v})
\end{equation}
This is approximated via a contrastive objective, which maximizes the log-likelihood of matched ("positive") samples while minimizing that of mismatched ("negative") samples:
\begin{equation}
	\mathcal{L}_{inter} \approx \log P_{\Theta}(\text{positive}) - \sum \log P_{\Theta}(\text{negative})
\end{equation}
For the bimodal (visual-textual) case, used when training on novel class data, this simplifies to:
\begin{equation}
	\begin{aligned}
		\mathcal{L}_{inter}(\Theta,\Phi) &= I(x_{v};x_{t}) \\
		&\approx \log P_{\Theta}(x_v, x_t) - \log\sum_{x'_{t}}P_{\Theta}(x_v, x'_{t}) \\
		&\quad - \log\sum_{x'_{v}}P_{\Theta}(x'_{v}, x_t)
	\end{aligned}
\end{equation}

\subsection{Overall Objective Function}
Our overall loss extends the multimodal ELBO from BraVL \cite{Du2023BraVL} by introducing weighted intra- and inter-modality mutual information regularization:
\begin{equation}
	\mathcal{L}(\Theta,\Phi,\Psi)=\mathcal{L}_{M}(\Theta,\Phi)+\lambda_{1}\mathcal{L}_{intra}(\Theta,\Phi,\Psi) +\lambda_{2}\mathcal{L}_{inter}(\Theta,\Phi)
\end{equation}

\subsection{Audio Modality Construction}
To capture auditory semantics and enable cross-modal alignment with EEG and textual representations, we construct an audio modality based on pretrained contrastive audio–language models.We employ the Contrastive Language–Audio Pretraining (CLAP) model (HTSAT-Unfused variant), a large-scale audio–language representation model trained to align auditory and textual semantics in a shared embedding space.All raw audio clips were loaded from the dataset directory and resampled to 48 kHz.Each audio waveform was processed using the CLAP processor to ensure consistent input normalization and tokenization.The model’s "getaudiofeatures function was used to obtain 512-dimensional latent representations.
To balance computational efficiency and memory usage, features were extracted in batches of 32 audio files and aggregated into a unified feature matrix.
The final feature set was stored in a structured .mat file for downstream multimodal learning.The resulting audio embeddings serve as the semantic representations for the auditory modality and are jointly optimized with other modalities under the multimodal ELBO and mutual information regularization framework.

\section{Experiments}
\subsection{Dataset}
Our experiments are conducted on the \textbf{THINGS-EEG} \cite{gifford2022large}dataset, which provides simultaneous brain and visual recordings for a large set of naturalistic object stimuli. To build our multimodal framework, we further constructed an additional \textbf{audio modality} derived from text-based image captions, resulting in a tri-modal dataset consisting of \textit{EEG}, \textit{Image}, and \textit{Audio} data.

\paragraph{EEG Modality.}
The EEG data and preprocessing pipeline used in this study follow the same configuration as in BraVL \cite{Du2023BraVL}. Specifically, we adopt the same subset of channels, filtering, and normalization strategies as described in that work to ensure methodological consistency and fair comparison. The EEG recordings originate from the THINGS-EEG dataset, which contains neural responses collected from human participants viewing images of 1,854 object categories. We selected a subset of \textbf{17 EEG channels} after preprocessing, including band-pass filtering and normalization, to remove artifacts and retain informative neural activity. Each EEG trial corresponds to a single image stimulus, providing a one-to-one mapping between brain and visual signals.

\paragraph{Image Modality.}
For the visual modality, we follow the same feature extraction process as BraVL \cite{Du2023BraVL}, where the \textbf{CORnet-S} model is used to preprocess the images and extract high-level visual representations. This model, inspired by the primate ventral visual stream hierarchy, provides biologically plausible feature embeddings that align well with cortical EEG activity patterns. By keeping the visual feature extraction consistent with prior work, we ensure that any performance difference can be attributed to the change from textual to auditory semantics rather than variations in visual representation.

\paragraph{Text Modality.}
Because BraVL only provides data after feature extraction, for comparison, I followed the original paper's process and used the BLIP model to generate a training set CSV of 16,540 entries and a test set CSV of 200 entries. After generating these CSVs, I extracted features (768 features) using the GPT-neo model used in the original paper, and then compared them with my subsequent audio modality content.

\paragraph{Audio Modality.}
In the original BraVL framework, the third modality is text, generated from image descriptions using the large-scale visual language model BLIP-Large. In our work, we propose replacing the text modality with an audio modality to capture richer perceptual cues more closely related to brain activity. Specifically, we generate a text-like mat dataset, but extracted from audio files, for comparative testing in this paper.

\paragraph{Data Alignment and Splitting.}
Each EEG sample, image, and audio segment corresponds to the same stimulus instance, thus forming a fully aligned trimodal dataset. We followed the original BraVL training and evaluation partitioning method.

\subsection{Tasks}
Our experimental evaluation is structured in a two-step process to rigorously assess the quality of the learned multimodal representations.

First, we train the BraVL model, as described in the Methods section, on the tri-modal dataset. The primary goal of this stage is to learn a shared latent space that effectively integrates information from EEG, image, and audio modalities.

Second, to validate the effectiveness of the learned representations, we perform a zero-shot classification task. We freeze the trained VAE encoder and use it to generate latent embeddings for the test data. A simple linear Support Vector Machine (SVM) classifier is then trained on these embeddings to predict the object category. This approach allows us to determine whether the latent space has successfully captured discriminative semantic information without overfitting to a complex downstream classifier.

We report both 50-way and 200-way classification accuracies to evaluate the generalization ability of the proposed BraVL model under different levels of semantic diversity. The 50-way task corresponds to novel-category decoding, while the 200-way task evaluates large-scale zero-shot decoding performance, providing a comprehensive test of the model's capabilities.

\subsection{Evaluation Metrics}
To quantify the performance of our model on the zero-shot classification tasks, we employ two standard evaluation metrics: Top-1 and Top-5 accuracy.

\begin{itemize}
	\item \textbf{Top-1 Accuracy:} This is a strict metric that measures the percentage of test samples for which the single class with the highest predicted probability is the correct class. It reflects the model's ability to pinpoint the exact correct category.
	\item \textbf{Top-5 Accuracy:} This metric measures the percentage of test samples for which the true class is among the five classes with the highest predicted probabilities. Top-5 accuracy is particularly useful for tasks with a large number of categories (e.g., 200-way classification), as it provides a more nuanced assessment of performance by giving credit to the model if the ground-truth label is considered plausible, even if it is not the top prediction.
\end{itemize}

Together, these two metrics provide a comprehensive and robust evaluation of the classification performance of the learned representations.

\subsection{Hyperparameters}
	The key hyperparameters for training the BraVL model on the EEG modality are summarized below:
	\begin{itemize}
		\item \textbf{Encoder/Decoder Architecture:} The encoders and decoders for all three modalities (EEG, vision, and audio) are implemented as two-layer Multi-Layer Perceptrons (MLPs) with 256 neurons in each hidden layer.
		\item \textbf{Latent Dimension:} The dimensionality of the shared latent space $z$ is set to 32.
		\item \textbf{Optimizer and Learning Rate:} We use the Adam optimizer with a fixed learning rate of $1 \times 10^{-4}$ to ensure stable training across modalities.
		\item \textbf{Training Schedule:} The model is trained for 500 epochs with a batch size of 1024. A larger batch size and longer training are employed to accommodate the scale of the EEG dataset.
		\item \textbf{MI Regularization:} The weighting parameters for the intra-modality and inter-modality mutual information regularizers, $\lambda_1$ and $\lambda_2$, are both set to 0.001. Experimental evidence indicates that the model performs optimally when these parameters are within the [0.001, 0.01] range.
		\item \textbf{Activation Function:} The ReLU activation function is used in the MLPs, consistent with standard deep learning practices.
		\item \textbf{SVM Classifier:} For the downstream classification task, a Support Vector Machine (SVM) with a Radial Basis Function (RBF) kernel is used. The `gamma` parameter for the kernel is set to $1 \times 10^{-3}$.
		\item \textbf{Implementation Framework:} The model is implemented using a combination of MindSpore and PyTorch.
    \end{itemize}

\section{Results}
This section presents the experimental results evaluating the performance of our proposed BraVL model across different modalities. We first establish a baseline by comparing our constructed audio modality against a standard text modality, followed by a multi-subject validation to assess the generalizability of our findings.

\begin{table*}[!t]
	\centering
	\caption{Performance comparison on the THINGS-EEG dataset.}
	\label{Tab:1}
	\begin{tabular}{llcccc}
	\toprule
	\multirow{2}{*}{\textbf{Subject}} & \multirow{2}{*}{\textbf{Modality}} & \multicolumn{2}{c}{\textbf{200-Way}} & \multicolumn{2}{c}{\textbf{50-Way}} \\
	\cmidrule(lr){3-4} \cmidrule(lr){5-6}
	& & \textbf{Top-1 Acc.} & \textbf{Top-5 Acc.} & \textbf{Top-1 Acc.} & \textbf{Top-5 Acc.} \\
	\midrule
	\multirow{2}{*}{Sub1} & English Text & 0.02269 & 0.08656 & 0.07000 & 0.26850 \\ %
	& Audio & 0.04169 & 0.14188 & 0.09825 & 0.32175 \\ %
	\addlinespace 
	\multirow{2}{*}{Sub2} & English Text & 0.02106 & 0.08581 & 0.06425 & 0.24725 \\ %
	& Audio & 0.03213 & 0.11338 & 0.08550 & 0.29100 \\ %
	\addlinespace
	\multirow{2}{*}{Sub3} & English Text & 0.02488 & 0.09075 & 0.06700 & 0.26700 \\ %
	& Audio & 0.03913 & 0.13875 & 0.10750 & 0.32600 \\ %
	\addlinespace
	\multirow{2}{*}{Sub4} & English Text & 0.01625 & 0.07419 & 0.04425 & 0.23250 \\ %
	& Audio & 0.03744 & 0.11550 & 0.09750 & 0.30250 \\ %
	\addlinespace
	\multirow{2}{*}{Sub5} & English Text & 0.01881 & 0.07475 & 0.05600 & 0.22875 \\ %
	& Audio & 0.02825 & 0.10175 & 0.08125 & 0.27375 \\ %
	\addlinespace
	\multirow{2}{*}{Sub6} & English Text & 0.02731 & 0.09756 & 0.07075 & 0.27250 \\ %
	& Audio & 0.04094 & 0.13675 & 0.09625 & 0.31850 \\ %
	\addlinespace
	\multirow{2}{*}{Sub7} & English Text & 0.02206 & 0.08650 & 0.06450 & 0.27025 \\ %
	& Audio & 0.05319 & 0.15825 & 0.13750 & 0.36000 \\ %
	\addlinespace
	\multirow{2}{*}{Sub8} & English Text & 0.03350 & 0.11094 & 0.09350 & 0.30300 \\ %
	& Audio & 0.05994 & 0.17169 & 0.14650 & 0.37475 \\ %
	\addlinespace
	\multirow{2}{*}{Sub9} & English Text & 0.01819 & 0.06988 & 0.04875 & 0.22150 \\ %
	& Audio & 0.03063 & 0.10469 & 0.07275 & 0.24750 \\ %
	\addlinespace
	\multirow{2}{*}{Sub10} & English Text & 0.02981 & 0.10713 & 0.08750 & 0.29950 \\ %
	& Audio & 0.04525 & 0.14313 & 0.11725 & 0.32225 \\ %
	\bottomrule					
	\end{tabular}
\end{table*}

\begin{figure}[!t]
	\centering
	\includegraphics[width=\columnwidth]{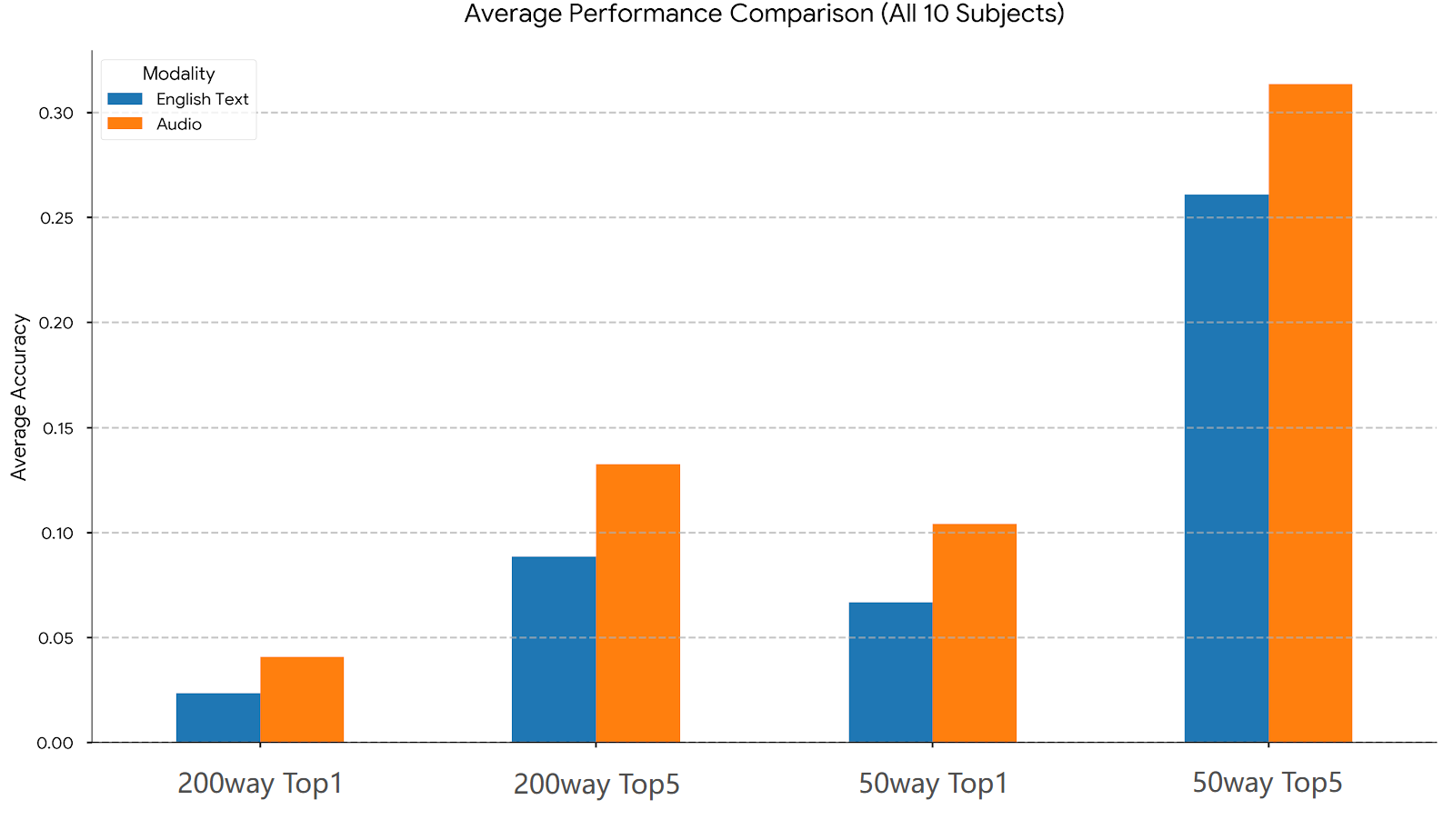} 
	\caption{Average performance comparison across different modalities.}
	\label{Fig:2}
\end{figure}

\subsection{Average classification accuracy}
\begin{table*}[h!]
	\centering
	\caption{Average classification accuracy across all 10 subjects.}
	\label{tab:average_performance}
	\begin{tabular}{lcccc}
		\toprule
		\textbf{Modality} & \textbf{200-Way Top-1} & \textbf{200-Way Top-5} & \textbf{50-Way Top-1} & \textbf{50-Way Top-5} \\
		\midrule
		English Text & 0.02346 & 0.08841 & 0.06665 & 0.26108 \\
		Audio & \textbf{0.04086} & \textbf{0.13258} & \textbf{0.10403} & \textbf{0.31380} \\
		\bottomrule
	\end{tabular}
\end{table*}

We first compared the mean zero-shot classification accuracy using the auditory modality and the text modality on 10 participants. As shown in Table~\ref{Tab:1} and Figure~\ref{Fig:2}, the mean accuracy of the ``Audio'' modality consistently outperformed the ``English Text'' modality across all four evaluation metrics.

Specifically:

\begin{itemize}
	\item In the most challenging 200-way Top-1 task, the mean accuracy of the auditory modality (0.04086) was significantly improved compared to the text modality (0.02346).
	
	\item In the 200-way Top-5 task, the accuracy of the auditory modality (0.13258) was also significantly higher than that of the text modality (0.08841).
	
	\item This advantage was also maintained in the 50-way task, with the auditory modality outperforming the text modality in both the Top-1 (0.10403 vs. 0.06665) and Top-5 (0.31380 vs. 0.26108) metrics.
\end{itemize}
\subsection{consistency across subjects}
\begin{figure}[!t]
	\centering
	\includegraphics[width=\columnwidth]{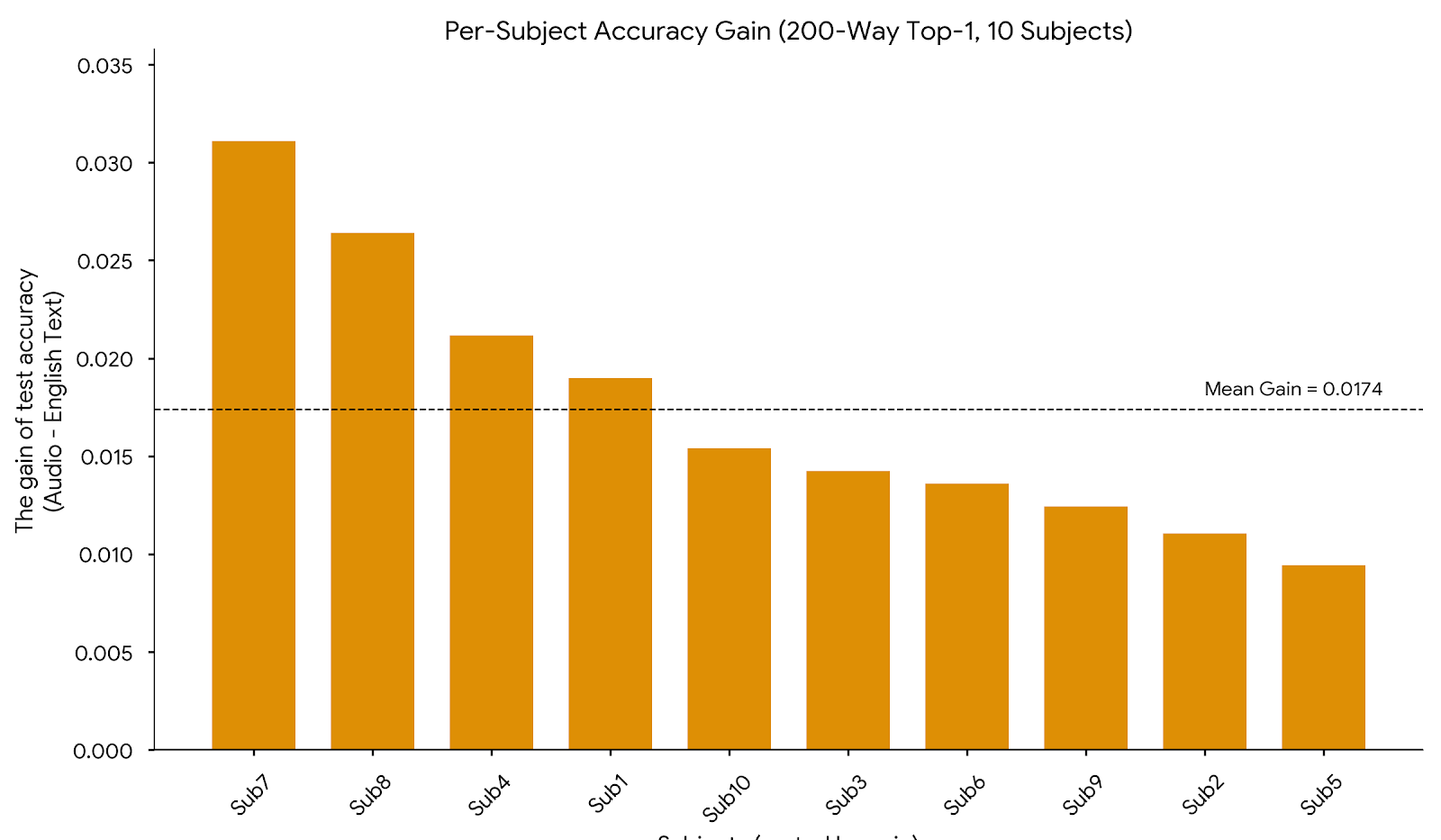} 
	\caption{Per-Subject Accuracy Gain (200-Way Top-1,10 Subjects)}
	\label{Fig:3}
\end{figure}

To assess the generalizability of this advantage, Figure~\ref{Fig:3} shows the accuracy gain (i.e., auditory modality accuracy minus text modality accuracy) for each participant in the 200-way Top-1 task.

The results clearly show that the gain was positive for all 10 participants. This finding indicates that the auditory modality advantage is not driven by an average effect across participants, but rather is a consistent phenomenon across individuals, although there are individual differences in the magnitude of the gain (e.g., between Sub7 and Sub5).

\begin{figure*}[!t]
	\centering
	\includegraphics[width=0.95\textwidth]{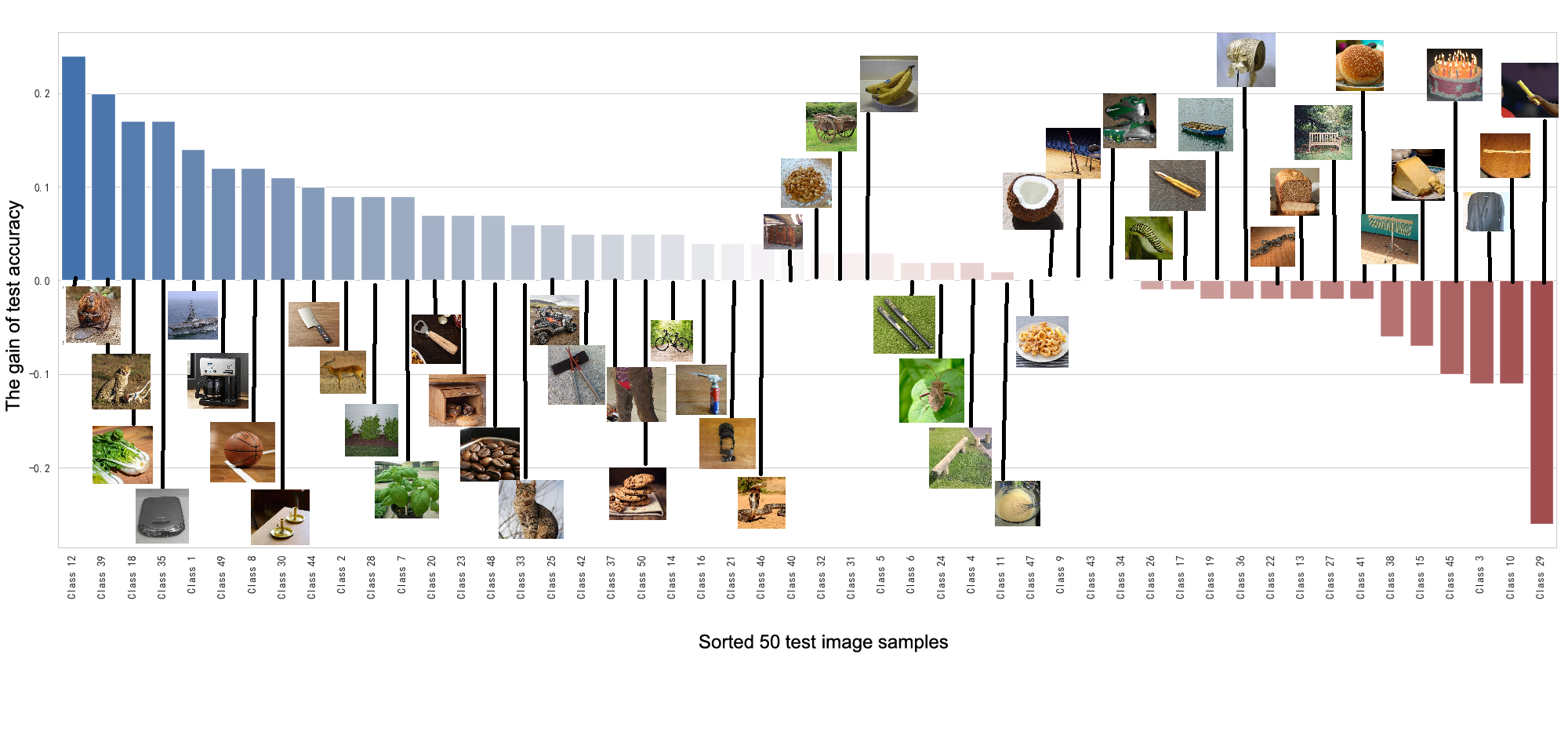}
	\caption{Gain of test accuracy across 50 individual test samples.}
	\label{Fig:4}
\end{figure*}
As further illustrated in Figure~\ref{Fig:4} across 50 test samples, the auditory modality exhibited a significant advantage in certain categories (indicated by the blue bars on the left), while the text modality performed better in a few categories (shown by the red bars on the right).
\subsection{Computational efficiency comparison}

\begin{table}[h!]
	\centering
	\caption{Computational efficiency comparison between modalities.}
	\label{tab:efficiency}
	\begin{tabular}{lcc}
		\toprule
		\textbf{Metric} & \textbf{English Text (GPT-NEO)} & \textbf{Audio (CLAP)} \\
		\midrule
		Training Time & 17.67 hours & 10.68 hours \\
		\addlinespace
		Embedding Dimension & 768 & 512 \\
		\bottomrule
	\end{tabular}
\end{table}
In addition to decoding accuracy, we also compared the computational efficiency of the two modalities (Table~\ref{tab:efficiency}).

\begin{itemize}
	\item \textbf{Feature Dimensions:} The CLAP-based auditory modality employs a more compact 512-dimensional feature set, compared to the 768 dimensions of the GPT-NEO-based text modality.
	\item \textbf{Training Time:} The total training time for the auditory modality (10.68 hours) is significantly shorter than for the text modality (17.67 hours), demonstrating a substantial efficiency advantage.
\end{itemize}

\section{Conclusion}
This study demonstrates that replacing textual representations with auditory modalities significantly enhances the performance of visual semantic decoding from brain signals. By introducing CLAP-based speech embeddings within a trimodal brain–vision–audio VAE framework, the proposed model achieves improved accuracy and computational efficiency, better reflecting the neural mechanisms of human language processing. These findings highlight the cognitive and practical advantages of auditory semantic representations and lay a foundation for developing more biologically plausible and efficient brain–language decoding systems.

\section{Discussion}
\subsection{Cognitive rationality of auditory representation}

These findings suggest that, compared to symbolic text input, speech signals exhibit greater consistency and ecological validity in aligning with brain activity patterns. In other words, auditory semantic representations are more closely aligned with the neural processing mechanisms underlying human language comprehension, thereby offering stronger physiological plausibility for modeling brain–semantic mappings.

This finding aligns with the dual-coding theory\cite{clark1991dual}, which emphasizes that human concepts rely on the collaborative encoding of the visual and language systems. Traditional text-based decoding models focus too much on the symbolic properties of language while neglecting the rich acoustic, prosodic, and other perceptual cues in speech. This study, by introducing an auditory modality, enables semantic representations not only to match brain activity at an abstract level, but more importantly, to better reflect the natural forms of human language perception, thus providing new evidence for revealing the interaction mechanisms between language and visual representations in the human brain.
\subsection{Methodological contributions and efficiency advantages}

Methodologically, this study, based on the BraVL framework, replaced text embeddings with speech embeddings extracted by the CLAP\cite{elizalde2023clap} model to construct a brain-visual-auditory joint variational autoencoder. Mutual information regularization further enhanced the potential semantic consistency across modalities.

The computational efficiency advantage observed in the Results (as shown in Table 3) is a key additional finding. The significant reduction in training time suggests that the auditory modality may provide a more direct mapping between brain signals and semantic representations, thereby reducing the computational complexity of feature matching. This efficiency improvement not only facilitates real-time decoding applications but also highlights the practical feasibility of auditory representations in future brain-computer interface systems.

\subsection{Limitations and Future Work}

Nevertheless, several limitations should be acknowledged. First, the auditory modality was generated using a text-to-speech (TTS) system; synthetic speech may differ from natural human speech in prosody and timbre, potentially affecting its correspondence with neural responses. Second, the current study utilized only English speech data, with limited availability of Chinese and other language datasets, leaving cross-linguistic generalizability to be further examined. Additionally, the present analysis relies on EEG data, which provides high temporal but limited spatial resolution, making it challenging to precisely delineate cortical mechanisms underlying auditory–visual integration. Future research could combine multi-modal neuroimaging methods such as fMRI or MEG to explore the neural foundations of multimodal semantic representations in both spatial and temporal domains.

Future work may also consider incorporating natural speech corpora (e.g., human reading or conversational speech) to improve ecological validity, investigating multilingual auditory modalities to assess cross-cultural consistency, and integrating higher-level acoustic attributes such as emotional tone and prosodic rhythm. Furthermore, leveraging large-scale pre-trained models (e.g., AudioCLIP, Whisper, or GPT-Audio series) could enable the construction of cognitively grounded brain–semantic alignment frameworks. We believe that the paradigm shift from text to speech not only enhances neural decoding performance but also contributes to the development of a more universal brain–language computational model aligned with human cognitive mechanisms, offering valuable theoretical support for the next generation of brain–computer interface systems.

\bibliographystyle{ieeetr}
\bibliography{references}

\vfill

\end{document}